\documentclass{ws-ijmpd}
\usepackage[super,compress]{cite}
\usepackage[colorlinks=true,allcolors=blue]{hyperref}
\newcommand{\ve}[1]{\mathbf{#1}}

\newcommand{\eref}[1] {Eq.~(\ref{#1})}
\newcommand{\tref}[1] {Table~\ref{#1}}
\newcommand{\fref}[1] {Fig.~\ref{#1}}

\begin{document}

\markboth{FABRIZIO DE MARCHI AND GIUSEPPE CONGEDO}
{SPACE TESTS	 OF THE STRONG EQUIVALENCE PRINCIPLE}

%
\catchline{}{}{}{}{}
%

\title{SPACE TESTS	 OF THE STRONG EQUIVALENCE PRINCIPLE: \\ BEPICOLOMBO AND THE SUN-EARTH LAGRANGIAN POINTS OPPORTUNITY
  }

\author{FABRIZIO DE MARCHI
}

\address{Department of Mechanical and Aerospace Engineering, Sapienza University of Rome, \\
Via Eudossiana, 18, 00184 Rome, Italy \\
\vspace{1pt}
fabrizio.demarchi@uniroma1.it
}

\author{GIUSEPPE CONGEDO}

\address{Institute for Astronomy, School of Physics and Astronomy, University of Edinburgh, \\
Royal Observatory, Blackford Hill, Edinburgh, EH9 3HJ, United Kingdom \\
\vspace{1pt}
Department of Physics, University of Oxford, \\
Keble Road, Oxford OX1 3RH, United Kingdom \\
\vspace{1pt}
giuseppe.congedo@ed.ac.uk}

\maketitle

\begin{history}
\received{Day Month Year}
\revised{Day Month Year}
\end{history}

\begin{abstract}
The validity of General Relativity, after 100 years, is supported by solid experimental evidence. However, there is a lot of interest in pushing the limits of precision by other experiments.
Here we focus our attention on the equivalence principle, in particular the strong form.
The results of ground experiments and lunar laser ranging have provided the best upper limit on the Nordtvedt parameter $\eta$ that models deviations from the strong equivalence principle. Its uncertainty is currently $\sigma[\eta]=  4.4 \times 10^{-4}$.
In the first part of this paper we will describe the experiment, to measure $\eta$, that will be done by the future  mission BepiColombo. The expected precision on $\eta$ is $\approx 10^{-5}$.
In the second part we will consider the ranging between the Earth and a spacecraft orbiting near the Sun-Earth Lagrangian points to get an independent measurement of $\eta$. In this case, we forecast a constraint similar to that achieved by lunar laser ranging.\\
\end{abstract}


\ccode{PACS numbers: 04.80.Cc, 95.30.Sf, 95.55.Pe, 96.30.Dz}




\section{Introduction}	

The equivalence principle (EP) \cite{misner73} states the equivalence between inertial and gravitational mass.
This fact is a mere coincidence in classical physics, but it has some important consequences, for example:
\begin{itemize}
\item the free fall of {\em any} object in the same gravity field depends only 
on their initial status and not on their composition or structure;
\item it is impossible to detect the difference between a uniform static 
gravitational field and a uniform acceleration: free-fall and inertial motion are physically equivalent.
\end{itemize}
As a consequence, the EP allows the geometrical description of spacetime, which is at the basis of  General Relativity (GR).

The weak form of the EP (WEP) is limited to strong and electroweak interactions. It can be verified by measuring the free fall of test masses with different
chemical compositions. Tests are performed on ground
with, for instance, torsion balances \cite{adelberger2009} or in space with low Earth
orbits (e.g.\ with the MICROSCOPE mission \cite{touboul2012}).\\
The strong form (SEP) extends the validity of the weak principle to self-graviting bodies. 
The EP violation for the body $i$ can be
parametrized as follows \cite{milani2002,damour1996}
\begin{equation}
m_i^G = m_i^I (1+ \delta_i + \eta \, \Omega_i),
\end{equation}
where $m_i^I$ ($m_i^G$) is the inertial (gravitational) mass, and
\begin{equation}
 \Omega_i=\frac{E_{g}}{m^{I}_i c^2}=-\frac{G}{2 m^{I}_i c^2}\iint \frac{  {d {m'}^\text{G}_i} {d {m''}^{G}_i}} {\vert \vert \mathbf r'-\mathbf r'' \vert \vert},
\end{equation}
where $c$ is the speed of light, and $E_g$ is the self-gravity energy, which is obtained by double-integrating over the mass of the body.
The WEP involves only the case $\Omega_i=0$ and corresponds to $\delta_i = 0$, while the SEP is valid, for each $\Omega_i$, when both $\delta_i$ and $\eta$ are equal to zero.

With experiments on ground, the typical $\Omega_i$ can be so small (see \tref{tab1}) that only the WEP can effectively be tested. 
The only means by which the SEP can be constrained is evidently in space involving celestial bodies.

\begin{table}[!h]
\tbl{Self-gravity coefficients $\Omega_i$ for some celestial bodies and a reference test mass.}
{\begin{tabular}{ll}
\toprule
 Sun  &  \hphantom{0} $-3.52 \times 10^{-6}$\\ 
 Jupiter &  \hphantom{0} $-1.21 \times 10^{-8}$\\ 
 Earth &  \hphantom{0} $-4.64 \times 10^{-10}$\\ 
 Moon &   \hphantom{0} $-1.88 \times 10^{-11}$\\ 
 test mass  (1~kg, size 5~cm)  &$\approx -8.90 \times 10^{-27}$\\
\botrule
\end{tabular} \label{tab1}
}
\end{table}

Thanks to retroreflectors placed on the facing side of the Moon it is possible to measure the Earth-Moon distance and detect a possible SEP violation signal.
This experiment was proposed by Nordtvedt \cite{nordtvedt1968}. In this case, a violation of the SEP will introduce a signal in the Earth-Moon range, its amplitude being proportional to $\Omega_\text{Earth}-\Omega_\text{Moon}$.

Over the last 46 years, the Lunar Laser Ranging (LLR) project has carried out a long sequence of range measurements, and the precision on the Earth-Moon relative differential accelerations is currently\cite{williams2009}
$\sigma[\delta a/a_\text{sun}] = 1.3 \times  10^{-13}$,  but this result includes possible violations of both SEP and WEP. Since ground experiments can test only the weak form of the EP, the parameter $\eta$ can be measured only by using the results of both experiments, ground and LLR. No disproofs of the SEP have still been found and the error associated to $\eta$ is currently\cite{williams2009} $\sigma[\eta]=4.4 \times 10^{-4}$. The BepiColombo mission is expected to improve this result by about an order of magnitude\cite{demarchi2016} -- a prediction is given in the first part of this paper. Instead, an alternative ranging experiment towards the Sun-Earth Lagrangian points -- recently proposed in Ref.\;\refcite{congedo2016} -- could easily reach the LLR's performance in a very short time span, which is investigated in the second part of the paper.

Therefore we will describe two experiments for the estimation of $\eta$. The first one in Section \ref{sect2} is the well-known Relativity experiment of the BepiColombo mission, while in Section \ref{sect3} we will study the same measurement performed on range data between the Earth and a spacecraft (SC) orbiting around a Sun-Earth Lagrangian point.

\section{MORE with BepiColombo}\label{sect2}
BepiColombo (BC) is a joint ESA/JAXA mission to Mercury with challenging objectives regarding geophysics, geodesy, and fundamental physics \cite{benkoff2010}. Currently, the launch is scheduled for the end of 2018, with a nominal duration of one year plus a possible one-year extension.

The Mercury Orbiter Radioscience Experiment (MORE)  is one of the on-board experiments that focus on gravimetry, rotation and Relativity \cite{milani2002, sanchez2006, cicalo2012}.
The goal is the measurement of key parameters by means of orbit determination techniques using the Earth-MPO \footnote{Mercury Planetary Orbiter.} radio link observables, i.e.\ range and range rate. The parameters for gravitation and rotation experiments are the Mercury gravity field coefficients, Love numbers, obliquity and libration. Instead the Relativity experiment consists in the measurement of the Parametrized Post-Newtonian (PPN) parameters, which account for possible small deviations from GR -- $\eta$ is one of them.

All parameters will be estimated by a global nonlinear least-squares fitting of all the {\em observed} signals (range, range-rate, accelerometer readings, etc.) along with the {\em computed} signals that are calculated by using mathematical models as accurate as possible. 
 The main characteristics of the Radioscience experiment are summarized in \tref{tab2}. For further details see Refs\;\refcite{milani2002,ashby2007}.
 The observed data of gravity and rotation experiments are primarily range-rate signals, which are poorly correlated with those of the Relativity experiment, i.e.\ Earth-MPO range only, because the frequency domains are very different. Since we are  interested in the Relativity experiment, we can neglect the motion of the MPO around Mercury (the orbital period is approximately 2~hrs)  and consider only the Mercury-Earth range.

\begin{table}[!h]
\tbl{Summary of the main characteristics of the radioscience experiments on-board BepiColombo.}
{\begin{tabular}{l l l l}
\toprule
                    & gravimetry                    &   rotation                                                       & Relativity\\
\hline
 parameters & -  gravity field coeffs      & - longitude libration                                    &- $\gamma, \beta, \alpha_1, \alpha_2,\eta$  \\
                     &      (up to the 25th deg.)          &     - obliquity                          &   - $\mu_0, \dot \mu_0 / \mu_0, J_{2 \odot}$ \\
                     &     - $k_2$                    &                                       &  - initial cond.\ of \\
                     &                                                                           &                                                                   & Earth and Mercury \\
observables  & range-rate                       & range-rate                              & range\\         
precision               &   $3.0\times 10^{-4}$ cm/s                               &   $3.0\times 10^{-4}$ cm/s      & $30$ cm @ $300$ s\\
                        &   @ $1000$ s                            &    @ $1000$ s                            &\\
freq.  domain   &   $\gtrsim 1.2\times 10^{-4}$ Hz &  $\gtrsim 1.2\times 10^{-4} $ Hz  &  $\approx 10^{-7}$ Hz\\
                 & (MPO mean motion)            &   (MPO mean   motion)     & (planetary mean motions)\\

\botrule
\end{tabular} \label{tab2}}
\end{table}

\subsection{Analytical model and sources of uncertainties}
We aim at calculating the expected root-mean-square (RMS) error of $\eta$ after the whole duration of the BC mission. Since the dare are obviously not available, we need to simulate them.
To this end, we are going present a simplified heliocentric analytical model that yields the perturbations on the Earth-Mercury range due to $\eta$ and all the parameters that are expected to correlate with. This is a typical Fisher/covariance analysis: the RMS of the parameters will be given by the square root of the diagonal elements of the covariance matrix.

We adopt the notation of  Ref.\;\refcite{moyer2003}: we define $\ve r_{ij} =\ve r_j - \ve r_i$
and $r_{ij} = \vert \vert \ve r_{ij} \vert \vert $, where $\ve r_i$ is the coordinate of the $i$th-body in an inertial reference frame. 
Planets are numbered from 1 (Mercury) to 8 (Neptune), while 0 refers to the Sun.
We also define the gravitational parameters for all bodies in the same way: $\mu_i = G m_i^G$.
The equations of motion for the $i$th-planet $i$, in the case $\eta \neq 0$, are
\cite{anderson1996,milani2002,turyshev2004,ashby2007, demarchi2016, congedo2016}
\begin{equation}
\label{eqr0i}
\ddot { \ve r}_{0i} = -\dfrac{\mu^\star}{r_{0i}^3}\ve r_{0i} + \displaystyle \sum_{j \neq i \neq 0} \mu_j \left[(1+ \eta \, \Omega_i)\dfrac{\ve r_{ij}}{r_{ij}^3} - (1+\eta\, \Omega_0) \dfrac{\ve r_{0j}}{r_{0j}^3}\right]  ,
\end{equation}
where the summation includes all solar system bodies (planets, dwarves planets, asteroids, etc.), and $\mu^\star= \mu_0+\mu_i+\eta(\mu_i \Omega_0+\mu_0 \Omega_i)$.
We can write a similar equation for body $k$ and afterwards calculate the range $\rho_{ik}=\vert \vert \ve r_{0i}-\ve r_{0k}\vert \vert$ where $i$ and $k$ are Earth and Mercury.
Since $\Omega_i \ll \Omega_0$ for all $i$,  the leading term is the last one, which is proportional to $\Omega_0$. It is an apparent term, essentially a perturbation on the acceleration of the Sun with respect to the Solar System Barycenter (SSB).
Note that there is a non-zero signal even if $\Omega_i=0$, which means that the experiment can be done also if the body $i$ is a drag-free test mass, e.g.\ a SC with an onboard accelerometer (see Section \ref{sect3}). 
It is worth mentioning that the signals due to other PPN parameters, such as $\beta, \gamma, \alpha_1,\alpha_2$, along with the effect due to $\zeta$ (the rate of change of $\mu_0$), $J_{2 \odot}$ (gravitational ``flattening'' of the Sun) and the initial conditions of Earth and Mercury (see Ref.\;\refcite{demarchi2016} for details), must all be calculated and included in the global fit. 
Also from \eref{eqr0i}, a high correlation among planetary perturbations (proportional to $\mu_j$s) and SEP violation is evident.

 In order to avoid systematic effects, the $\mu_j$s must be added to the set of parameters to be estimated, and their errors must be taken into account in terms of prior constraints in the global covariance analysis.
Current uncertainties of planetary $\mu_j$s range from $2.8\times 10^{-4}$ (Mars) to $10.5$~km$^3$/s$^2$  (Neptune) \cite{luzum2011}. 
Regarding asteroids, their relative errors can be very large (50\% or more).

 To summarize, we will calculate the signatures on the  Earth-Mercury range due to all the following effects:
 \begin{enumerate}
  \item initial conditions of Earth and Mercury;
  \item SEP violation -- free parameter: $\eta$;
 \item planets/dwarf planets/asteroids -- free parameters: $\mu_j$;
 \item secular variation of the Sun's gravitational parameter $\mu_0$ -- free parameters: $\delta_{\mu_0}$ (bias of the measured $\mu_0$ from the true value at the starting epoch), and its rate of change in time $\zeta=\dot \mu_0/\mu_0$,
 \item PPN -- free parameter: $\bar \beta=\beta-1$,
 \item Sun's quadrupole coefficient: free parameter $J_{2 \odot}$, whereas higher order terms are negligible.
\end{enumerate}
 The PPN parameter $\gamma$, which is related to the curvature produced
 by unit rest mass, has not been considered here for simplicity. 
 However, this is not reductive since the best estimate of $\gamma$ ($\sigma[\gamma]=2.0\times 10^{-6}$) 
is expected to be given right after the dedicated superior conjunction experiment
 (SCE) during the cruise phase of BC.
The value of the Nordtvedt parameter can be derived from the Nordtvedt quation 
\begin{equation}
\label{eq:nordt}
\eta=4 \beta - \gamma -3,
\end{equation}
which will be used as a prior.
We also neglect the preferred frame parameters $\alpha_1$ and $\alpha_2$ since they are poorly correlated with the other parameters of the Relativity experiment, in particular $\eta$. For more details compare the results of experiments A, B, C and D in Ref.\;\refcite{milani2002}.
Finally, we assume that the unperturbed orbits of planets and asteroids are circular with radius $R_{0i}$, and co-planar.
We define $\ve q$ as the vector of all $N_p$ parameters, $q_m \delta \ve r_{i,m}$ is the displacement from the circular reference orbit $\ve R_i=R_{0i}\ve u_r^i$ for the $i$th-body due to the (linearized) force $q_m \delta \ve f_{i,m}$ relative to the (small) parameter $q_m$.

The procedure is as follows:
\begin{enumerate}
\item write the  heliocentric position of the $i$th-body as 
\begin{equation}
\ve r_i= \ve R_i+\sum_{n=1}^{N_p} q_n \delta \ve  r_{i,n};
\end{equation}
\item for each $q_m$, decompose $\delta \ve r_{i,m}$ and the perturbative force $\delta \ve f_{i,m}$ into radial, along-track and out-of-plane components
\begin{equation}
\begin{split}
 \delta \ve r_{i,m}&=x_i \ve u_r^i+y_i \ve u_t^i+z_i \ve u_w^i, \\
 \delta \ve f_{i,m}&=R_m^i \ve u_r^i+T_m^i \ve u_t^i+W_m^i \ve u_w^i ;
 \end{split}
\end{equation}
\item solve the Hill's equations for  $i=1$ and $i=3$
\begin{equation}
\label{eq:hill}
\begin{split}
\ddot x_i - 2\,n_i \dot y_i -3 \,n_i^2 x_i &=R_m^i ,\\
\ddot y_i+2 \,n_i \dot x_i  &= T_m^i ,\\
\ddot z_i+ n_i^2 \dot z_i  &= W_m^i ,
 \end{split}
\end{equation}
where $n_i$ is the mean motion of the $i$th-body;
\item finally calculate the Earth-Mercury range as
\begin{equation}
 \rho_{13}(t,\ve q)= \vert \vert \ve r_{13} \vert \vert \approx R_{13}+ \sum_n q_n \frac{\delta \ve r_{13,n}  \cdot \ve R_{13}}{R_{13}}
\label{eq:pert1}
 \end{equation}
where $\delta \ve r_{13,n}=\delta \ve r_{3,n}-\delta \ve r_{1,n}$ and
the factor $1/R_{13}$ can be rewritten in Legendre polynomials $P_n$
\begin{equation}
\frac{1}{R_{13}}= \frac{1}{R_{03}}\sum_{l=0}^\infty \left( \frac{R_{01}}{R_{03}}\right)^l P_l (\cos \Phi_{13}),
\end{equation}
where $\Phi_{ij}=(n_j-n_i)\,t+\varphi_j-\varphi_i$.
\end{enumerate}

Due to visibility windows, range and range-rate data contain several gaps. A gap
occurs approximately every day and lasts about 9.3 h.
A low-frequency sampling ($f_s =10^{-4}$ Hz) is therefore sufficient for our purposes since the
involved signals have frequencies of the same order of
planetary mean motions. We can then calculate the range at epochs $t_i$ and obtain the  looked-after $N_p\times N_p$
 Fisher matrix, or \textit{normal matrix}. Including all prior information, it is given by
\begin{equation}
F_{jk} = \sum_{i=1}^{N}  \frac{1}{\sigma_i^2} \frac{\partial \rho_{13} (t_i,\ve q_0)}{\partial q_j} \frac{\partial \rho_{13} 
(t_i,\ve q_0)}{\partial q_k} + \frac{1}{2}\frac{\partial^2 P(\ve q)}{\partial q_j \partial q_k} ,
\label{eq:fisher}
\end{equation}
where $N$ is the number of range measurements;
$\sigma_i$ is the RMS error on each data point\footnote{For the Ka-band we adopted $\sigma_i = 15 \sqrt{300 f_s} \mbox{ cm }= 2.6$ cm \cite{milani2010}.};
$P(\ve q)$ is a function that contains all prior information (the Nordtvedt equation \eref{eq:nordt} and the uncertainties on all the $\mu_m$s) and is given by
\begin{equation}
P(\ve q) =\frac{(\eta-4 \bar \beta)^2}{\sigma_N^2}+\sum_m  \frac{(\mu_m- \mu_m^P)^2}{\sigma_{\mu_m}^2};
\end{equation}
$\mu_m^P$ are the measured values of $\mu_m$ and $\sigma_{\mu_m}$ are the corresponding errors;
the summation over $m$ is extended to all $GM$s;
$\sigma_N=2.0 \times 10^{-6}$ is the expected RMS error of $\gamma$ after the expected performance of the SCE.
The inverse of $F_{jk}$ yields the covariance matrix, whose diagonal elements give us the expected RMS errors, and correlations, of all the parameters.

\subsection{Results}
As well as standard parameters, we include the $\mu_j$s of all the
 planets and the 343 more massive asteroids (the total 
number of parameters was 362). 
Since some of the $\mu_j$s  are expected to be improved by GAIA \cite{mouret2009} and JUICE, we calculate the global covariance by 
 using the expected RMS errors of $\mu_j$ at the epoch of the mission.
The RMS error of all parameters, including the initial conditions of Mercury and Earth, are reported in \tref{tab:relnom}. 
Regarding the SEP violation,  we found $\sigma[\eta]=3.13 \times 10^{-5}$. If we were to compare this result with the ``idealistic case''  where the $\mu_j$s have all zero errors \cite{milani2002,schettino2015,cicalo2016}, we would
 find that the uncertainties 
degrade the precision of most of the PPN parameters by about an order of
magnitude. However, since the current RMS error of $\eta$, from
LLR measurements, is $\sigma[\eta]=4.4 \times 10^{-4}$, we can conclude that
the BC Relativity experiment will improve the
current constraint on $\eta$ by a factor of 10 at least, having included uncertainties on the planetary masses.

\begin{table}[!h]
\tbl{Expected formal errors for the Relativity experiment on-board BepiColombo.}
{\begin{tabular}{lll}
\toprule
 parameter                                  &   units                                 &           RMS error        \\
 \hline               
 $\beta$                         &     -                               &  $    7.81 \times 10^{-6 }$        \\                       
 $\eta$                           &     -                               &  $ {\bf 3.13 \times 10^{-5 }}$    \\              
 $\mu_0$                       &      [cm$^3$s$^{-2}$]     & $  5.50 \times 10^{13 }$          \\                      
 $J_{2\odot}$                &      -                               &  $  8.03 \times 10^{-10}$          \\                       
$\zeta=\dot \mu_0/\mu_0$  &     [yr$^{-1}$]                 &  $  1.78 \times 10^{-14}$   \\ 
$X_{1}$                       &     [cm]                          &   $  2.49 \times10^3$               \\ 
 $Y_{1}$                       &     [cm]                          &  $   1.18 \times10^4$              \\                                                        
 $Z_{1}$                       &     [cm]                          &   $  5.15	$                         \\                                                         
 $\dot X_{1}$                &     [cm s$^{-1}$]            &    $ 2.36 \times 10^{-3 }$         \\                              
 $\dot Y_{1}$                &     [cm s$^{-1}$]            &    $ 1.68 \times 10^{-3 }$         \\                         
 $\dot Z_{1}$                 &     [cm s$^{-1}$]            &   $  4.72 \times 10^{-6 }$         \\                            
 $\dot X_{3}$                 &     [cm s$^{-1}$]            &   $  1.77 \times 10^{-3 }$         \\                          
 $\dot Y_{3}$                 &     [cm s$^{-1}$]            &  $  9.41 \times 10^{-5 }$           \\ \botrule
\end{tabular}
}
 \label{tab:relnom}
\end{table}

\section{An opportunity with the Lagrangian points}\label{sect3}

When testing for a SEP violation, the advantage of the ranging between 
 two planets over that between Earth and Moon is twofold: a longer 
baseline  ($\approx 1$ vs $\approx 3\times10^{-3}$ AU) and 
 $\delta a/a_\text{sun} \propto \Omega_0$ instead of $\Omega_\text{earth}-\Omega_\text{moon}$. 
This in turn implies a much bigger ranging signal amplitude (about three orders 
of magnitudes better than the Nordtvedt effect \cite{turyshev2004,milani2009}).
In fact, even if the time span and the precision of the data will be worse, 
 a bigger self-energy and a stronger signal will certainly allow better 
measurements of $\eta$. For example, consider the BC experiment: the expected measurement precision on the SEP is $\sigma[\delta a/a_\text{sun}] \approx 10^{-11}$, 
which will be roughly two orders of magnitude worse than WEP measurements achieved by LLR 
and torsion balances experiments\cite{adelberger2009}. 
However, since the signal is $\propto \Omega_0$,  the parameter $\eta$ will be constrained with an accuracy of $10^{-5}\text{--}10^{-6}$ (see Section \ref{sect2} and also Ref.\;\refcite{milani2002}), which is of course better than LLR. This is also the case of the Lagrangian points ranging, with the only difference that a smaller baseline will give us an RMS error that will be similar in magnitude to LLR.

\begin{figure}[h!]
\centerline{\psfig{file=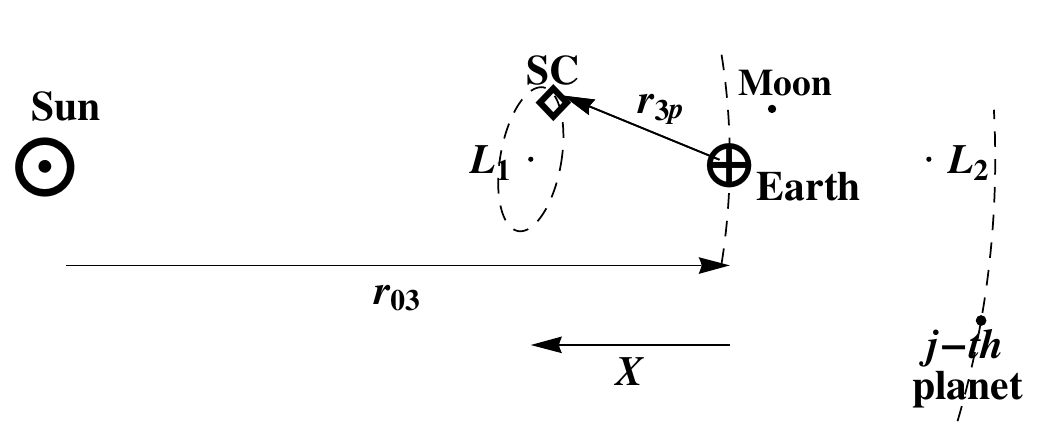,width=0.59\columnwidth}}
\caption{Spacecraft ranging towards $L_1$ or $L_2$ as a means by which to test the SEP (not in scale). We calculate the SEP signature as a perturbation on the Earth's orbit around the the Sun (${\bf r}_{03}$) as well as on the SC ranging (${\bf r}_{3p}$). We also include perturbations from other planets. \label{f1}}
\end{figure}

\subsection{Detailed calculations}
\begin{figure}[h!]
\centerline{\psfig{file=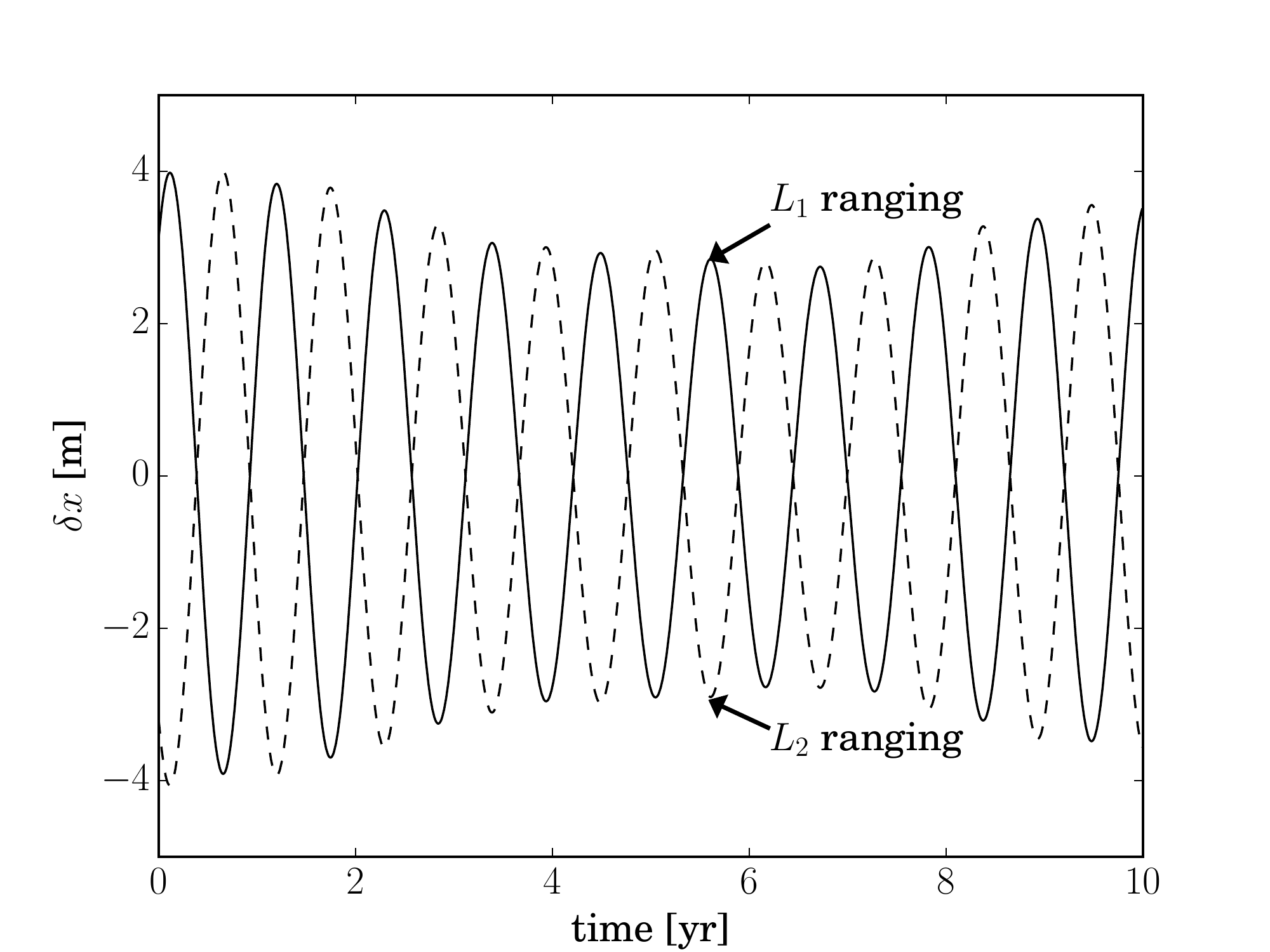,width=0.59\columnwidth}}
\caption{Range perturbations (normalised to $\eta=1$) for a SC orbiting either $L_1$ or $L_2$. \label{f3}}
\end{figure}

In the Earth's reference frame, the positions of the collinear Lagrangian points are the solutions of the following equation
\begin{equation}\label{eq:L1}
-\frac{\mu_0}{ \vert R-X \vert^3} (R - X) + \mu_3 \left(\frac{X}{ \vert X \vert ^3} -\frac{1}
{R^2}\right)+  n_3 ^2 (R-X) =0,
\end{equation}
where $R$ is the Eart-Sun distance, and $n_3$ is the mean motion of the Earth.
 \eref{eq:L1} has three solutions: $X_{1,2}\approx\pm 0.01$ AU that correspond to $L_1$ and $L_2$, 
and $X_3\approx 2$ AU that corresponds to $L_3$.
We will consider only the case of $L_1$ and $L_2$ as these are the spots where many missions fly to.
Consider a SC, hereafter identified with the index $p$,  near $L_1$ (or $L_2$). Its mass and self-gravity 
energy are negligible with respect to those of the Sun and all planets.
The SC's equation of motion relative to the Sun can be obtained by \eref{eqr0i} after this substitution: $(\Omega_3,\mu_3,\ve r_{03},\ve r_{3j}) \rightarrow (0,0,\ve r_{0p},\ve r_{pj})$.
We subtract the SC's equation of motion 
from \eref{eqr0i} to finally derive the 
relative motion, $\ve r_{3p}$, between the SC and Earth, which is given by
\begin{equation}\label{eq:L1_rel}
\ddot {\ve r}_{3p} = - \mu_0 \left( \dfrac{\ve r_{0p}}{r_{0p}^3}-  \dfrac{\ve r_{03}}{r_{03}^3}\right) 
-\mu_3 \dfrac{\ve r_{3p}}{r_{3p}^3} + \sum_{j\neq 0,3} \mu_j \left(\dfrac{\ve r_{pj}}{r_{pj}^3}-\dfrac{\ve r_{3j}}{r_{3j}^3}\right)+ \eta\, 
\Omega_3 \sum_{j \neq 3} \mu_j \dfrac{\ve r_{j3}}{r_{j3}^3},\\
\end{equation}
where $\ve r_{0p}=\ve r_{03}+\ve r_{3p}$. It is worth noting that we are in fact solving the equation of motion for the observed SC ranging, $\ve r_{3p}$.
As it was done for the Earth-Mercury range in the previous section, we decompose $\ve r_{3p}=\{\delta x, \delta y\}$ in radial and along-track components (but now only $\delta x$ can be measured).
For simplicity we assume that the SC is very near to the Lagrangian point, such that the gravity field can be linearized in this case, and all trajectories are Lissajous orbits.
 Details of the calculation can be found in Ref.\;\refcite{congedo2016}. In \fref{f3} we plot $\delta x$ (normalised to $\eta=1$) for the two scenarios of a SC orbiting around either $L_1$ or $L_2$.

In order to compute our prediction for a measurement of the SEP around the Lagrangian point, we assume we have $N$ equally-spaced observations of the SC's range distance, over a total observation of $T=5$ yr, sampling interval $\delta t= 1$ h\footnote{Hereafter we assume an hour integration time for all range measurements.}.
We can then calculate the Fisher matrix from \eref{eq:fisher}. The free parameters considered in our analysis are: $\eta$, the initial position and velocity of the Earth and the initial position and velocity of the SC.
We distinguish between two possible scenarios.
In the \textit{realistic scenario} (A) we use a nominal range error typical for two-way ranging in the X-band, $\sigma_i=0.1$ m
\footnote{As obtained from a degradation of a conservative factor 2.5 of the Ka-band range error 
$\sigma_i=0.15 \sqrt{300/\delta t}\approx0.04$ m\cite{iess2001,schettino2015,cicalo2016}, owing to the lower frequencies typical of the X-band.}.
Additionally, we assume the following prior uncertainties on the orbital initial conditions: 
\begin{enumerate}
\item 2 m and $3\times10^{-5}$ m/s for the Earth's heliocentric radial position and velocity, 
from a great abundance of radio tracking data 
\cite{kaplan2015};
\item 145 m for the Earth's heliocentric along-track position as this is less well constrained \cite{kaplan2015}; 
\item no assumed prior on both the Earth's heliocentric along-track velocity as this is very weakly constrained by current data, and the parameters of the SC's orbit relative to Earth. 
\end{enumerate}
In the \textit{optimistic scenario} (B) we use the range error
typical of the Ka-band, $\sigma_i=0.04$ m, 
as well as a factor 10 improvement in the knowledge of the Earth's initial position and velocity, 0.2 m and $3\times10^{-6}$ m/s, which is likely to be achieved in the near future.

\subsection{Results}
Neglecting errors in planetary masses and ephemerides, we forecast $\sigma[\eta]=  6.4(2.0) \times 10^{-4}$ (5 yr integration time) via Earth-$L_1$ ranging  in a realistic (optimistic) scenario depending on current (future)
range capabilities and knowledge of the Earth's ephemerides. A combined measurement, $L_1 + L_2$, gives instead an improved constraint of
$4.8(1.7) \times 10^{-4}$,
which would be comparable with those already achieved by LLR.
It is worth noting that the performances could be much improved if data were integrated over time and over the number of satellites
flying around either of the two Lagrangian points. We point out that some systematics (gravitational
perturbations of other planets or figure effects) are much more in control compared to other experiments.
This SC ranging would be a new and complementary probe to constrain
the strong equivalence principle in space.

\subsection{Conclusions}
In this work we described two experiments devoted to testing the SEP in space. In both cases we performed a global covariance analysis based on simulated data.

The first test is the BC Relativity experiment: we calculated the effect of the uncertainties on the masses of the Solar System's bodies on the estimation of PPN parameters. 
We forecast a degradation for the RMSs of all parameters, including $\eta$ for the strong equivalence principle, of about an order of magnitude with respect to the nominal case where uncertainties are not taken into account. Nonetheless this result, in terms of $\eta$, represents an improvement of a factor 10 over the current precision achieved by LLR.

In the second part of the paper we calculated the signal due to SEP violation on the ranging between a ground station and a SC orbiting near an Earth-Sun collinear Lagrangian point. 
With a covariance analysis based on a 5 years mission, we forecast an RMS error for $\eta$ that would be around the same level of current measurements by LLR and ground experiments.
We conclude that this recently proposed experiment would serve as a direct test of the SEP that is both independent from other experiments, and at least comparable in terms of performances achieved in a relatively short time span.

\section*{Acknowledgments}
FDM acknowledges the advice and support of the Celestial Mechanics group of Pisa. GC acknowledges support from Hertford College, Harding Fund, 
the Beecroft Institute for Particle Astrophysics and Cosmology, and Oxford Martin School.  
The results of the research presented in the first part of this work
have been performed within the scope of Contract No. ASI/
2007/I/082/06/0 with the Italian Space Agency.

\end{document}